\documentclass{PoS}
\rightline{\parbox{4cm}{\large\rm
ADP-19-2/T1082\\
DESY 19-017\\
LTH 1196
}}

\usepackage{graphicx,amsmath,amssymb}
\usepackage{fontenc, times, mathptmx}
\usepackage{braket}
\usepackage{subcaption}
\usepackage{wrapfig}

\newcommand{\dlm}{\delta m}

\newcommand{\latSmall}{32^3 \times 64}
\newcommand{\latLarge}{48^3 \times 96}

\title{Anomalous magnetic moment of the muon with dynamical QCD+QED}
\ShortTitle{Anomalous Magnetic Moment of the Muon}

\author{A.~Westin$^{a}$, R.~Horsley$^{b}$, W.~Kamleh$^{a}$, Y.~Nakamura$^{c}$,
	 H. Perlt$^{d}$, P.~E.~L.~Rakow$^{e}$, G.~Schierholz$^{f}$ 
 	 A. Schiller$^{d}$, H. St\"uben$^{g}$, R.~D.~Young$^{a}$,
  \speaker{J.~M.~Zanotti}$^{,a}$\\
	\llap{$^a$} CSSM, Department of Physics, The
        University of Adelaide, Adelaide SA 5005, Australia\\
        \llap{$^b$} School of Physics and Astronomy,
                    University of Edinburgh,
                    Edinburgh EH9 3FD, UK \\
        \llap{$^c$} RIKEN Advanced Institute for Computational
                    Science, Kobe, Hyogo 650-0047, Japan \\
        \llap{$^d$} Institut f\"ur Theoretische Physik,
                    Universit\"at Leipzig, 04103 Leipzig, Germany \\
        \llap{$^e$} Theoretical Physics Division,
                    Department of Mathematical Sciences,
                    University of Liverpool,
                    Liverpool L69 3BX, UK \\
        \llap{$^f$} Deutsches Elektronen-Synchrotron DESY,
                    22603 Hamburg, Germany \\
        \llap{$^g$} RRZ, Univeristy of Hamburg, 20146 Hamburg, Germany\\
        E-mail:\\
        \email{alex.westin@adelaide.edu.au, \\
          james.zanotti@adelaide.edu.au}}

\author{CSSM/QCDSF/UKQCD Collaboration}

\abstract{
  The current $3.5\sigma$ discrepancy between experimental
  and Standard Model determinations of the anomalous magnetic moment
  of the muon $a_\mu=(g-2)/2$ can only be extended to the discovery
  $5\sigma$ regime through a reduction of both experimental and
  theoretical uncertainties.
  On the theory side, this means a determination of the hadronic vacuum
  polarisation (HVP) contribution to better than 0.5\%, a level of precision
  that demands the inclusion of QCD + QED effects to properly understand
  how the behaviour of quarks are modified when their electric charges
  are turned on.
  The QCDSF collaboration has generated an ensemble of
  configurations with dynamical QCD and QED fields with the specific
  aim of studying flavour breaking effects arising from differences in
  the quark masses and charges in physical quantities.
  Here we study these effects in a calculation of HVP around the SU(3)
  symmetric point.
  Furthermore, by performing partially-quenched simulations we are
  able to cover a larger range of quark masses and charges on these
  configurations and then fit the results to an SU(3) flavour breaking
  expansion.
  Subsequently, this allows for an extrapolation to the physical point.}

\FullConference{The 36th Annual International Symposium on Lattice
                Field Theory - LATTICE2018\\
                22-28 July, 2018\\
                Michigan State University, East Lansing, Michigan, USA.}

\begin{document}
\section{Introduction}
\label{sec:intro}
There currently exists a \( 3.5-4\) standard deviation discrepancy
between the experimentally measured anomalous magnetic moment of the
muon, $a_\mu = \frac{g-2}{2}$, and current Standard Model predictions
(see e.g. \cite{keshavarziNewAnalysis}).
At present, the experimental \cite{Bennett:2006fi} uncertainty and the
total theoretical uncertainties are of comparable magnitude.
The planned Muon $g-2$ Experiment at Fermilab \cite{E989} aims to
reduce the experimental uncertainty to 140 parts-ber-billion.
Thus it is essential to get the theoretical uncertainties down to a
comparable precision --- this will require the ``hadronic vacuum
polarisation'' (HVP) contributions to be known to better than $0.5\%$.
Simulations of the QCD-only contribution to HVP have received a surge
of interest over the past few years, with results now being quoted at
the physical point with errors ${\cal O}(1\%)$.
At this level of precision, contributions from QED effects are
expected to play a role.
In this talk, we present preliminary results for the electromagnetic
contributions to the hadronic vacuum polarisation tensor --- the key
ingredient relevant to the QCD contribution to $(g-2)_\mu$.

\section{Accessing \(a_\mu^{HVP}\)}
\label{sec:access}
We explore two methods of extracting the HVP contribution to the
anomalous magnetic moment from the lattice.
First we will take a more traditional approach where we determine the
vacuum polarisation from the vacuum polarisation tensor, $\Pi_{\mu\nu}
\left(Q \right) $, as first described in
\cite{Blum:2002ii,Gockeler:2003cw}.
Secondly we will investigate the time-momentum representation
method proposed in \cite{Bernecker:2011gh} to extract a value for \(
a_\mu^{HVP} \).

\subsection{Vacuum polarisation tensor}
\label{sec:vpt}
We can calculate \(a_\mu^{HVP}\) from the vacuum polarisation
function $\Pi(Q^2)$ using
\begin{equation}
  \label{eq:amu}
  a_\mu^{HVP} = 4\alpha^2 \int^\infty_0 dQ^2 K(Q^2; m^2_\mu) \left\{
  \Pi(Q^2) - \Pi(0) \right\},
\end{equation}
where $K(Q^2; m^2_\mu)$ is a known kernel function \cite{Blum:2002ii},
and the polarisation function $\Pi(Q^2)$ is determined from the
polarisation tensor
\begin{equation}
  \label{eq:vacPolTen}
  \Pi_{\mu\nu} = \int d^4x e^{iQ\cdot x} \left< J_\mu(x) J_\nu(0)
  \right> = \left( Q_\mu Q_\nu - \delta_{\mu\nu}Q^2 \right)\Pi
  \left(Q^2\right).
\end{equation}

\subsection{Time-moment representation}

In the time-moment representation (TMR), the vacuum subtracted
polarisation function, $\hat{\Pi}(Q^2) \equiv 4\pi^2\left(
\Pi(Q^2) - \Pi(0) \right)$, is obtained from the
spatially summed two-point correlator, \( G(t) \),
\begin{align}
  \hat{\Pi}(Q^2) &= 4\pi^2 \int_0^\infty dt G(t) \left( t^2 -
  \frac{4}{Q^2}\sin^2\left(\frac{Qt}{2}\right)\right), \\ 
  G(t) &= -\int d^3x \left< J_i(x) J_i(0) \right>.
\end{align}
Substituting this into Eq.~(\ref{eq:amu}), one finds
\begin{equation}
  \label{eq:timeMom}
  a_\mu^{HVP} = \left( \frac{\alpha}{\pi} \right)^2 \int_0^\infty dt
  G(t) \tilde{K}(t;m_\mu),
\end{equation}
where 
we employ the analytic form for \(\tilde{K}(t;m_\mu)\) derived in
\cite{DellaMorte:2017dyu}.

\section{Simulation details}
We follow the flavour-breaking program outlined
in~\cite{Bietenholz:2010jr,Bietenholz:2011qq} originally for QCD, and
extended to include electromagnetic interactions
in~\cite{Horsley:2015vla,Horsley:2015eaa}.
Starting from the symmetric point $m_u = m_d = m_s$, our strategy is
to keep the singlet quark mass $\bar{m}=(m_u+m_d+m_s)/3$ fixed at its
physical value while $\delta m_q \equiv m_q-\bar{m}$ is varied.
This procedure leads to highly constrained polynomials in $\delta m_q$
and $e_q^2$, and thus reduces the number of free parameters
drastically.

For the partially-quenched, flavour-diagonal $a\bar{a}$ ($a=u, d, s$)
octet (vector) meson masses, with all annihilation channels turned
off, a group theoretical analysis incorporating both QCD and QED terms
leads to the mass formula to leading order in $\alpha_{\rm EM}$ and
second order in $\delta m_q$
\begin{align}
  \label{eq:psMass}
  M(a\bar{a}) = &M_0 + 2\alpha\delta\mu_a + \beta_0\frac{1}{6} (\delta
  m_u^2 + \delta m_d^2 + \delta m_s^2) + 2\beta_1\delta\mu_a^2 +
  \beta_0^{EM}(e_u^2 + e_d^2 + e_s^2) \nonumber \\
  & + 2\beta_1^{EM}e_a^2 + \gamma_0^{EM}(e_u^2\delta m_u + e_d^2\delta
  m_d + e_s^2\delta m_s) + 2\gamma_1^{EM}e_a^2\delta\mu_a \nonumber
  \\
  & + 2\gamma_4^{EM}(e_u^2 + e_d^2 + e_s^2)\delta\mu_a +
  2\gamma_5^{EM}e_a(e_u\delta m_u + e_d\delta m_d + e_s\delta m_s)\ .
\end{align}
We have distinguished between sea ($m_q$) and valence
(partially-quenched, PQ) quark masses $\mu_a$ with $\delta
\mu_a=\mu_a-\bar{m}$.

The introduction of quark charges complicates the definition of
an SU(3) symmetric point.
In~\cite{Horsley:2015vla} we introduced the Dashen scheme which
absorbs all electromagnetic effects in the neutral, purely connected
pseudoscalar mesons $(m_\pi^{a\bar{a}})$ into the definition of the
quark mass, which we refer to as the ``Dashen mass'' $\mu^D_a$.
This drastically simplifies the flavour-breaking expansions of the
pseudoscalar mesons \cite{Horsley:2015vla}, while the effect on the
expansion of the vector mesons as needed here is to replace
$\delta\mu_a$ in Eq.~(\ref{eq:psMass}) with $\delta\mu_a^D$.
A natural definition for the SU(3) symmetric point in this scheme is
then one where \( m_\pi^{u\bar{u}} = m_\pi^{d\bar{d}} = m_\pi^{s\bar{s}} \).
This tuning was performed on two volumes in \cite{Horsley:2015vla}.

We employ five ensembles of fully dynamical QCD+QED lattice
configuations generated by the QCDSF collaboration, including
simulations on two different volumes, \( \latSmall \), and \(
\latLarge \), with lattice spacing $a= 0.068(1)$fm, and an
exaggerated QED coupling $\alpha_{\rm EM}\sim 0.1$.
Our simulation set-up employs the so-called QED$_L$ formulation
\cite{Hayakawa:2008an}, where the zero mode of the photon field is
removed on each time slice for the valence quarks.
However, since in this work we only consider electrically neutral
$q\bar q$ hadronic systems, photon zero modes are in any case unlikely
to have any effect.

The details of the five ensembles are summarised in
Table~\ref{table:ens} where we provide the masses of the unitary
neutral and charged pseudoscalar mesons.
In order to better constrain the coefficients of the flavour-breaking
expansions, on each ensemble we
employ partially-quenched quark masses corresponding to neutral
pseudoscalar meson masses in the range \( 260 \leq
m_{q\bar{q}} \leq 770\)\,MeV.
Quark charges are also partially quenched in that we allow for charges
\( Q_q \in \left( 0, -\frac{1}{3\sqrt{13}}, +\frac{2}{3\sqrt{13}}, \pm
\frac{1}{3}, \pm \frac{\sqrt{2}}{3}, \pm \frac{2}{3} \right) e \).
At our enhanced QED coupling ($e\approx
\sqrt{13.7}\,e^{\text{phys}}$), the quark charges
$Q_q=\left( -\frac{1}{3\sqrt{13}}, +\frac{2}{3\sqrt{13}}\right) e$ allow for
simulations to be performed with near-physical valence quark charges.

\begin{table}
  \center
  \begin{tabular}{l |  l l l l l l l l}
      Ensemble &   $L^3\times T$ &  \(N_f\) & \(m_{u\bar{u}}\) &
      \(m_{d\bar{d}}\) & \(m_{s\bar{s}}\)  & \(m_{q\bar{q}}^{min}L\) &
      \(m_{\pi^+}\) & \( m_{K^+} \) \\
      \hline \hline
      1   & $\latSmall$ &     2+1  &     430       &       405      &      405       &       4.4    &     435     &   435 \\
      2   & $\latSmall$ &     2+1  &     360       &       435      &      435       &       4.0    &     415     &   415 \\
      3   & $\latSmall$ &   1+1+1  &     290       &       300      &      570       &       3.2    &     320     &   470 \\
      4   & $\latLarge$ &     2+1  &     430       &       405      &      405       &       6.7    &     435     &   435 \\
      5   & $\latLarge$ &     2+1  &     360       &       435      &      435       &       5.9    &     420     &   420
  \end{tabular}
  \caption{Ensembles used in this work. All masses are in MeV.}
  \label{table:ens}
\end{table}

As first observed in \cite{Boyle:2017gzv}, we find a clear charge
dependence of the vector current renormalisation constant, \(Z_V\).
This will be discussed in more detail in a forthcoming publication.

\section{Results and discussion}

\subsection{Finite Volume Effects}
\label{sec:fve}
When working on a finite four-torus with dimensions $L^3\times T$, the
single polarisation function in Eq.~(\ref{eq:vacPolTen}), as valid for
$O(4)$, is replaced by five independent functions corresponding to the
five irreducible representations of the finite cubic symmetry group
$H(3)$ \cite{Bernecker:2011gh,Aubin:2015rzx}
\begin{equation}
  \label{eq:irrep}
\begin{array}{rlrl}
  A_1 &: \sum\nolimits_i \bar{\Pi}_{ii} = \left( 3q^2 - \vec{q}^{\,2} \right) \bar{\Pi}_{A_1}, &
  \quad T_1 &: \bar{\Pi}_{4i} = -\left(q_4q_i\right) \bar{\Pi}_{T_1} \\
  A_1^{44} &: \bar{\Pi}_{44} = \left( \vec{q}^{\,2} \right) \bar{\Pi}_{A_1^{44}}, &
  \quad T_2 &: \bar{\Pi}_{ij} = -\left(q_iq_j\right) \bar{\Pi}_{T_2}, i \neq j, \\
  & & \quad E &: \bar{\Pi}_{ii} - \sum\nolimits_i\bar{\Pi}_{ii}/3 = \left(
  -q_i^2 + \vec{q}^{\,2}/3 \right) \bar{\Pi}_E.
\end{array}
\end{equation}

\begin{figure}
  \begin{subfigure}[t]{0.47\textwidth}
    \includegraphics[width=\textwidth]{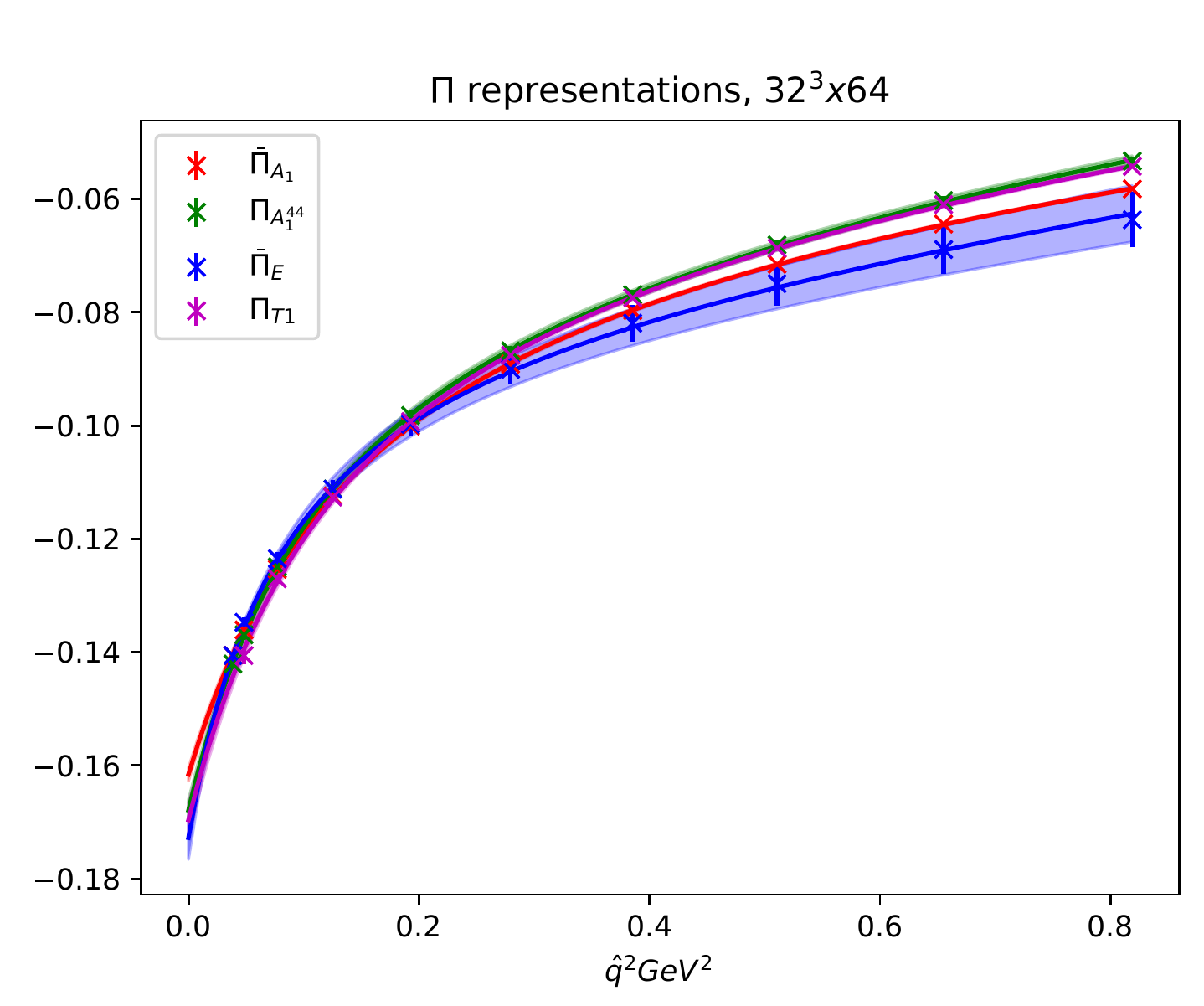}
  \end{subfigure}
  \hspace{0.06\textwidth}
  \begin{subfigure}[t]{0.47\textwidth}
    \includegraphics[width=\textwidth]{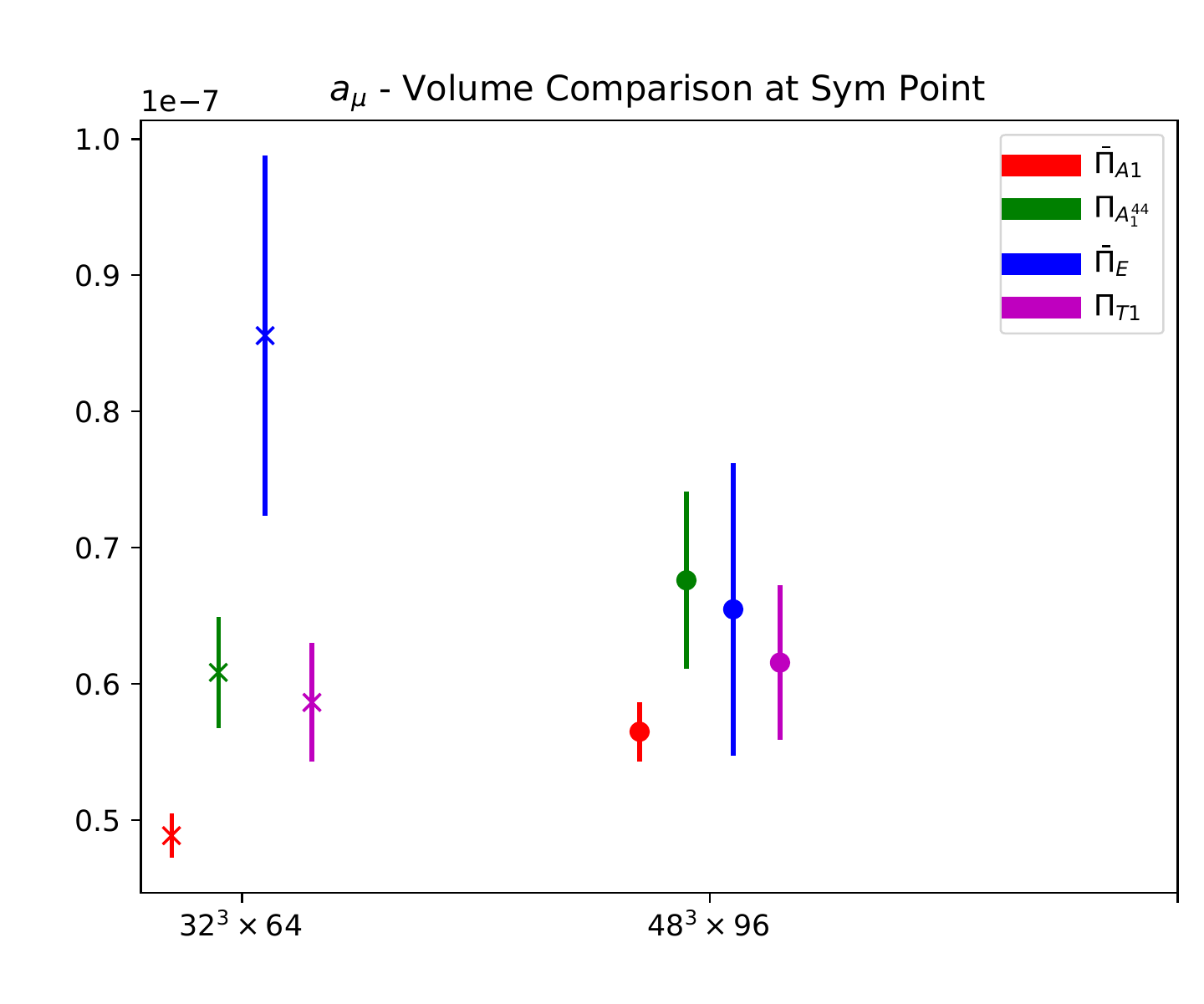}
  \label{fig:vacPol:volume}
  \end{subfigure}
\vspace{-5mm}
  \caption{Left: Polarisation functions from the $A_1,\,
    A_1^{44},\,E,\,T_1$ irreducible representations from the
    $32^3\times 64$, ensemble 1. Right: $a_\mu^{HVP}$ determined via
    the vacuum polarisation functions belonging to the different
    irreducible representations of the finite volume for two volumes
    (ensembles 1 and 2).}
  \label{fig:vacPol}
\end{figure}

These five functions should agree in the infinite volume and continuum
limits, hence we are provided with a method for investigating the
impact of the finite volume on our results.
In left plot of Fig.~\ref{fig:vacPol}, we display the $A_1,\,
A_1^{44},\,E,\,T_1$ polarisation functions obtained from the
$32^3\times 64$ volume close to the SU(3) symmetric point (i.e. ensemble 1
in Table~\ref{table:ens}).
Here we observe a clear discrepancy between the irreducible
representations of the vacuum polarisation tensor and indicates the
presence of finite volume effects in the simulations.

This behaviour is carried through to $a_\mu^{HVP}$ after we follow the
procedure outlined in Sec.~\ref{sec:vpt}.
This is seen by the scatter of the data points displayed in the right
plot of Fig.~\ref{fig:vacPol} for the $\latSmall$ volume.
When we repeat the process for the larger $\latLarge$ volume at the
same quark masses (ensemble 2), we observe a pronounced reduction in the
scatter of results obtained from the different irreducible
representations.
This provides us with confidence that results obtained on the larger
volume have only a small remnant finite size effect.

\subsection{Time Moment}
\begin{figure}
  \begin{subfigure}[t]{0.47\textwidth}
    \includegraphics[width=\textwidth]{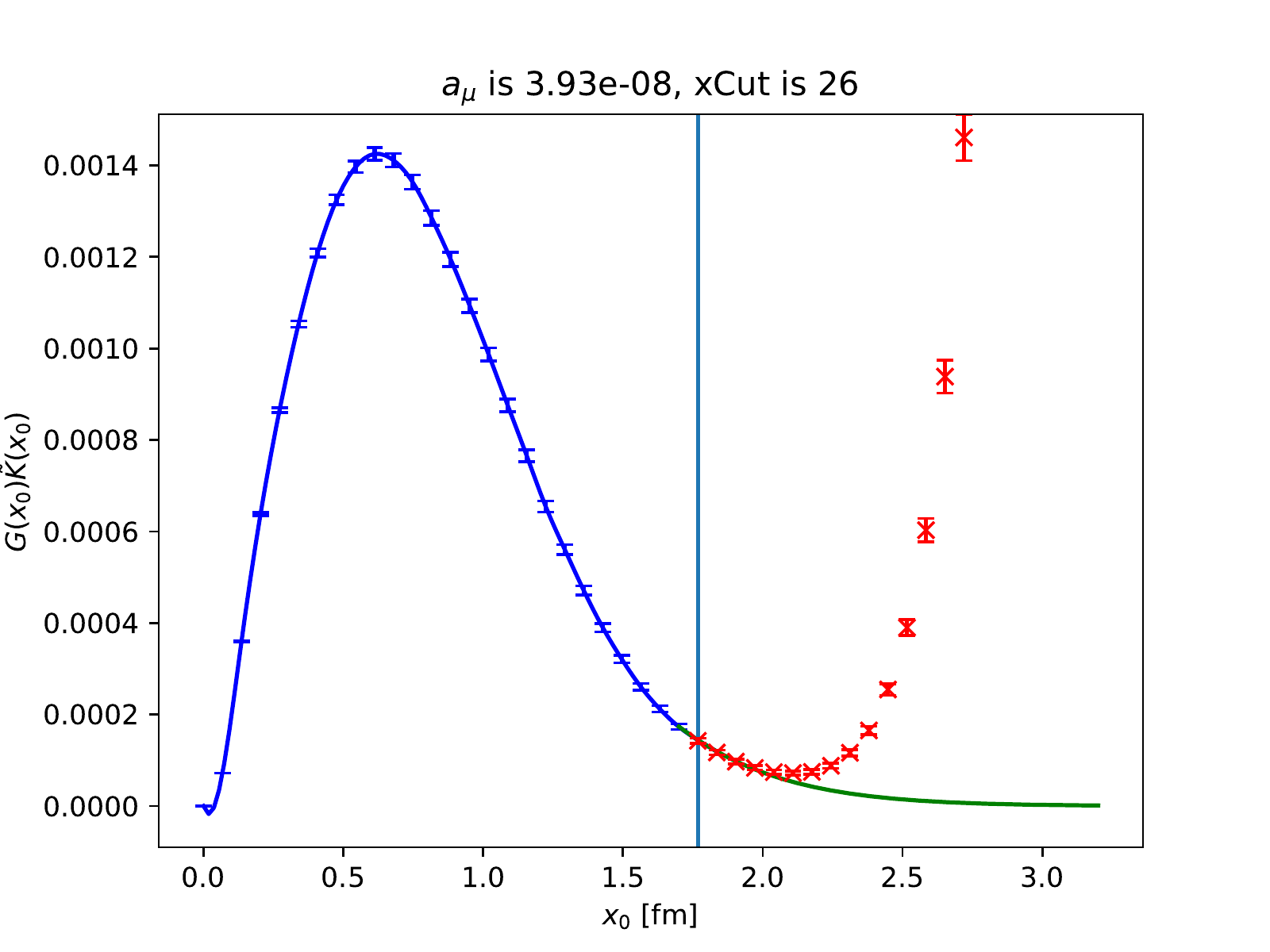}
    \caption{Blue points are correlator data used in constraining a
      cubic spline (blue curve).
      Red crosses are correlator data after the \(t_{cut}\)
      (vertical line), which are ignored. Green line in the tail is
      from our exponential function for \(t > t_{cut} \)
      region. \label{fig:tcut}}
  \end{subfigure}
  \hspace{0.06\textwidth}
  \begin{subfigure}[t]{0.47\textwidth}
    \includegraphics[width=\textwidth]{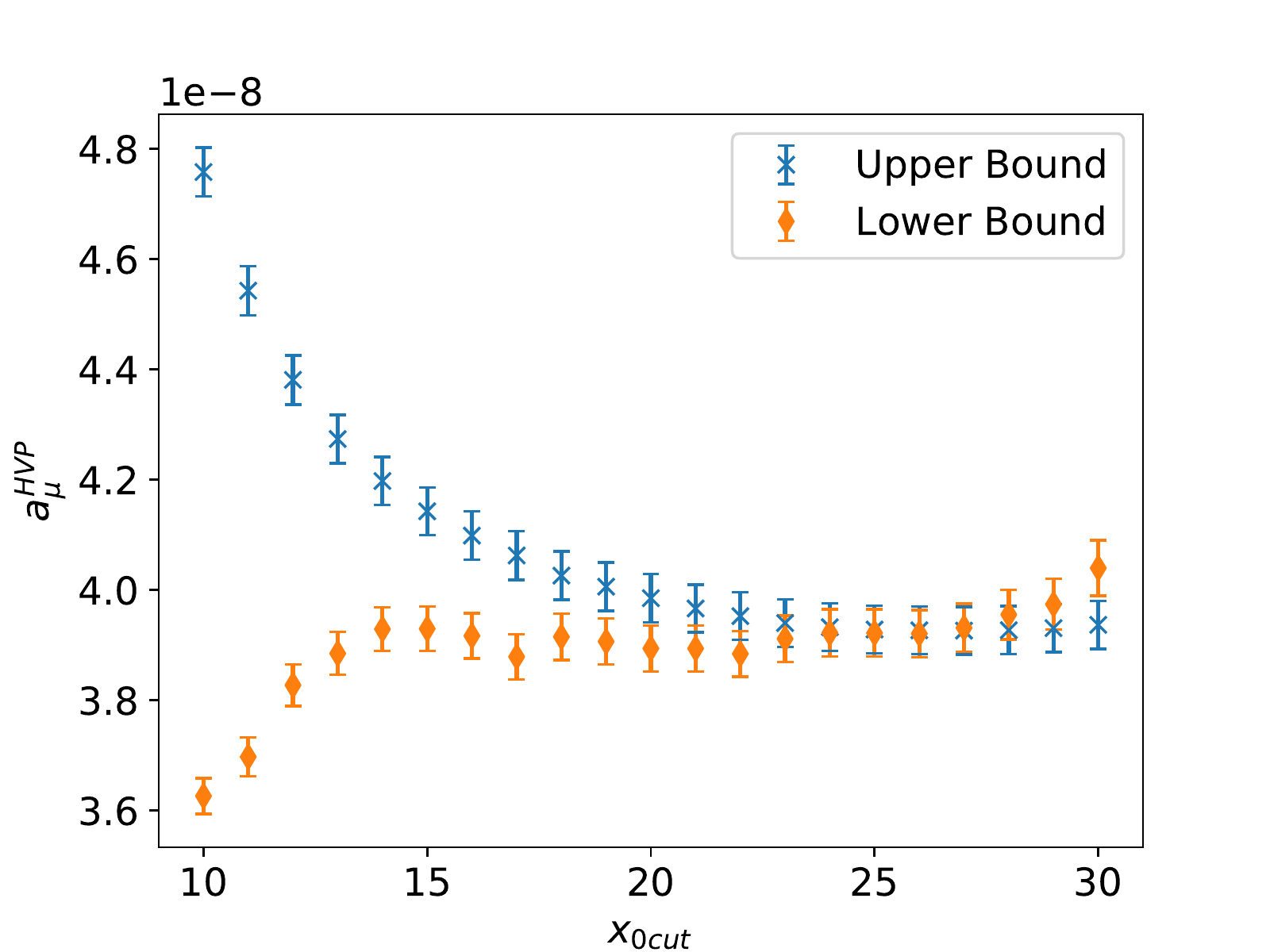}
    \caption{Bounding method \cite{Borsanyi:2016lpl}. Upper and lower bounds
      agree around the \(t = 26\) mark, which verifies our choice of
      \(t_{cut} = 26\) for this particular correlator.\label{fig:bound}}
  \end{subfigure}
  \caption{}
  \label{fig:xcut}
\end{figure}

We will now turn our attention to determining $a_\mu^{HVP}$ from the
time moment representation as given in Eq.~(\ref{eq:timeMom}),
following the method proposed in \cite{DellaMorte:2017dyu}.
At large times, the 2 point function \(G(t)\) suffers from a loss of
signal into statistical noise and is contaminated by the backwards
propagating state.
Since Eq.~(\ref{eq:timeMom}) requires $G(t)$ to be known to infinite
times, this issue is overcome by only using the 2 point function data,
$G^{\text{data}}(t)$ up to some value of \(t=t_{cut}\).
After this time, we fit a single exponential with the ground state
vector meson mass, $E_0$, such that
\begin{equation}
  G(t) = \left\{ \begin{array}{lr}
    G^{\text{data}}(t)     & t \leq t_{cut}, \\
    Ae^{-E_0 t} & t > t_{cut}.
  \end{array} \right.
\end{equation}
For the region \(t < t_{cut} \) we use a cubic spline over the lattice
data before computing the contribution of this region to the integral
in Eq.~(\ref{eq:timeMom}).
We choose \( t_{cut} \) such that the single exponential ansatz
matches the data before the signal is lost to noise, and that it forms
a smooth continuous line with the spline of that data at \(t_{cut}\).
An example for $t_{cut}=26$ on ensemble 1 is shown in
Fig.~\ref{fig:tcut}.

We can check our choice of \(t_{cut}\) using the recent bounding
method \cite{Borsanyi:2016lpl,Blum:2018mom}.
For this we define \(G(t)\) as
\begin{equation}
  G(t) = \left\{ \begin{array}{lr}
    G^{\text{data}}(t)     & t \leq t_{cut}, \\
    G^{\text{data}}(t_{cut})e^{-E(t-t_{cut})} & t > t_{cut},
  \end{array} \right.
\end{equation}
where we have an upper bound from \( E = E_0 \) and a lower bound from
\( E = \log \left[ \frac{G(t_{cut})}{G(t_{cut} + 1)} \right] \). 
When these two bounds agree, we find the optimal choice for $t_{cut}$.

In Fig.~\ref{fig:bound} we see that the upper and lower bounds
converge at \( t_{cut} = 26 \), which matches with when our
exponential fit matches on smoothly with \( G(t) \) in
Fig.~\ref{fig:tcut}.

We note that this procedure can easily be improved by including states
beyond the ground state, allowing for smaller values of $t_{cut}$ to
be used \cite{Borsanyi:2016lpl,Blum:2018mom}.
This will be pursued in future work.

\begin{figure}
  \begin{subfigure}[t]{0.5\textwidth}
    \includegraphics[width=\textwidth]{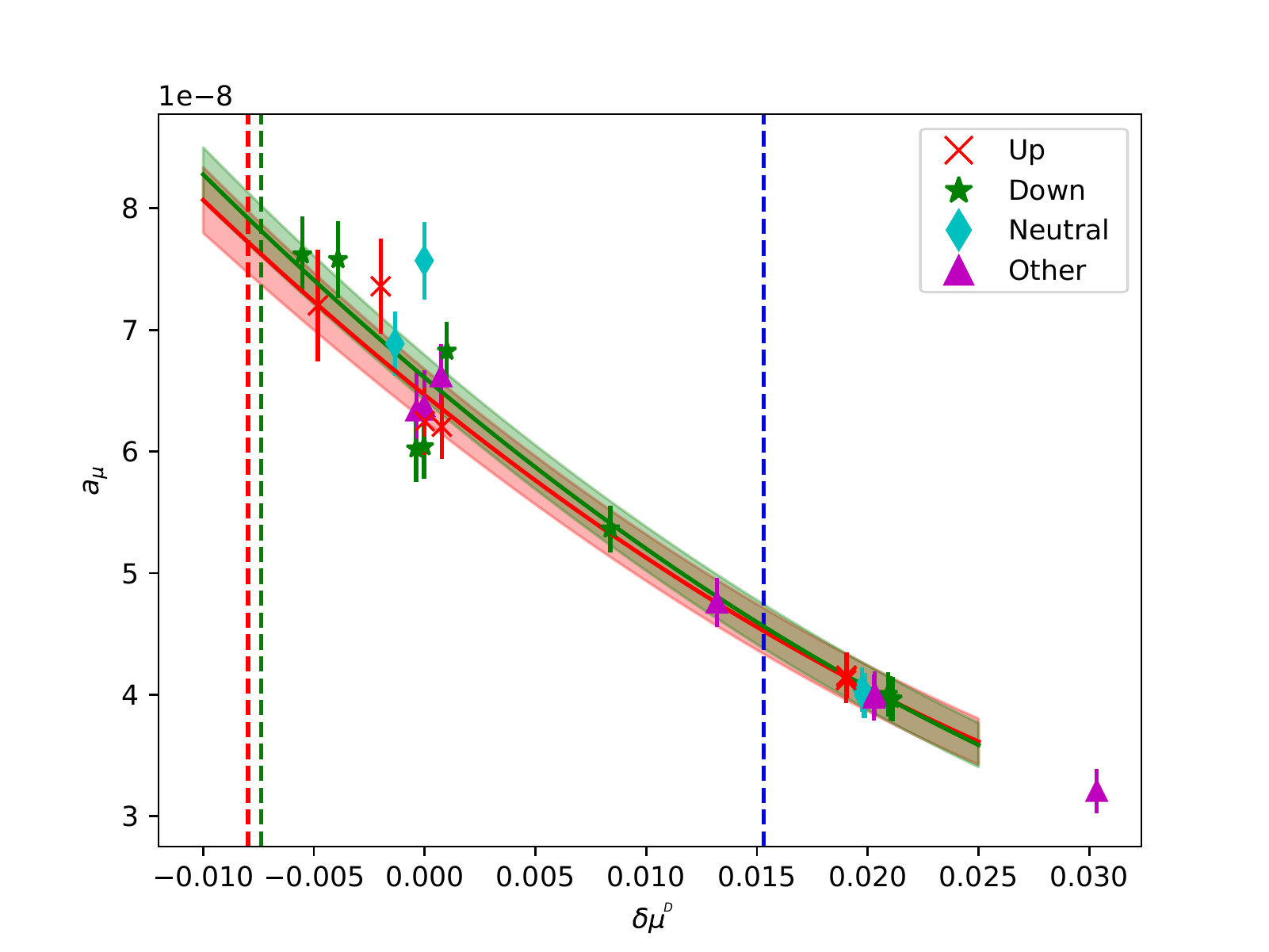}
  \end{subfigure}
  \begin{subfigure}[t]{0.5\textwidth}
    \includegraphics[width=\textwidth]{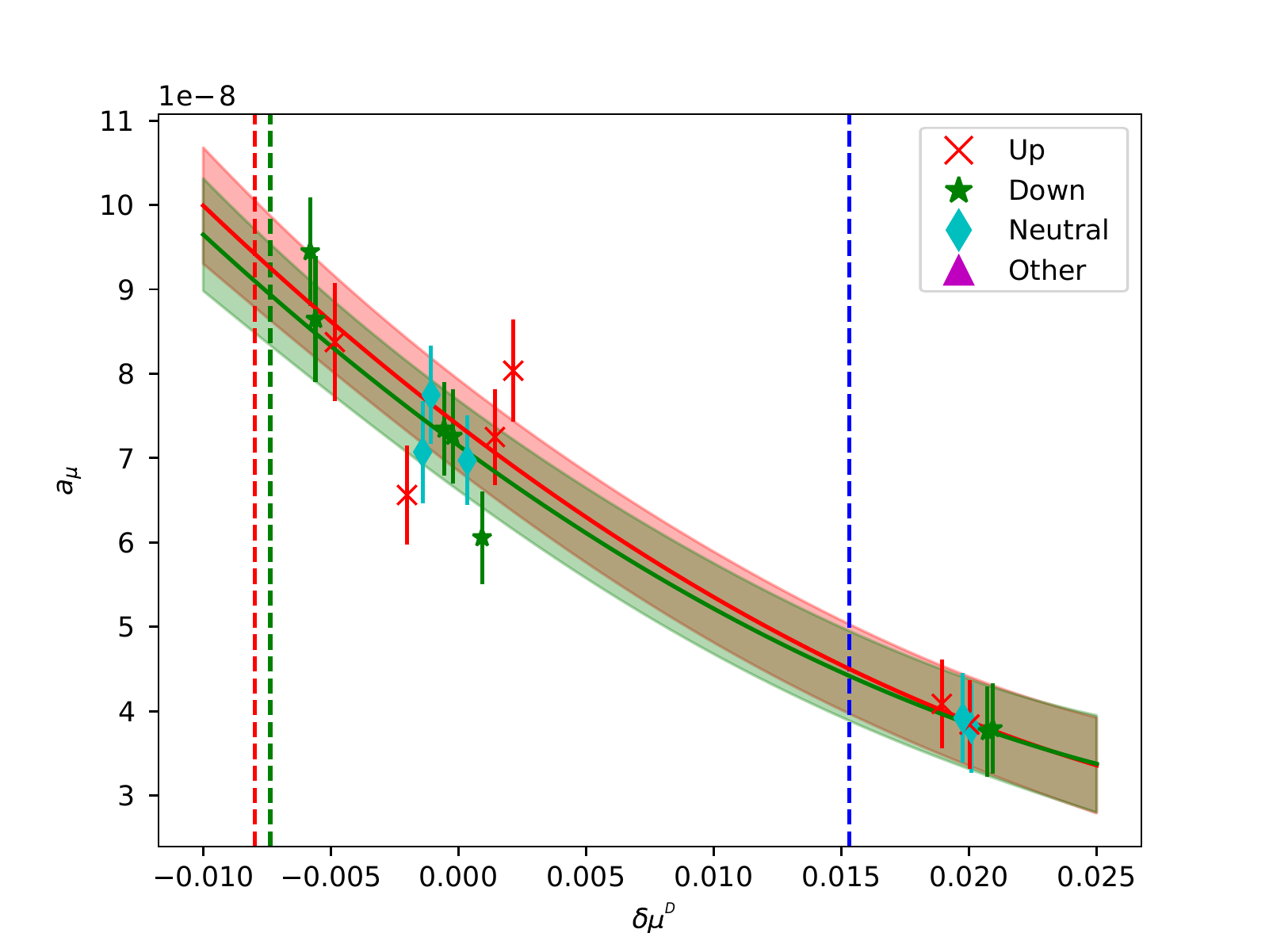}
  \end{subfigure}
  \caption{\(a_\mu^{HVP}\) against Dashen mass, \emph{left}: \(
    \latSmall \), \emph{right}: \( \latLarge \). Note that points are
    shifted to \(\dlm = \dlm^{phys}\) line. Colours refer to quarks
    with different charges, \emph{red}: Up quark, \emph{green}:
    Down/strange quarks, \emph{cyan}: `Neutral' quark, \emph{magenta}:
    other charges.}
  \label{fig:32Extrapolate}
\end{figure}

The above bounding method is then repeated for all partially quenched
quarks on all five ensembles in Table~\ref{table:ens}.
We can then calculate \(a_\mu^{HVP}\) on each of our ensembles.
These are plotted against the Dashen mass in
Fig.~\ref{fig:32Extrapolate} for $\latSmall$ (left) and $\latLarge$
(right) volumes.
Recalling the flavour-breaking expansion for the flavour-diagonal
vector mesons given in Eq.~(\ref{eq:psMass}), then since the
SU(3)-flavour properties of $a_\mu^{HVP}$ are the same, we can apply
the same expansion for \(a_\mu^{HVP}\) to extrapolate to the physical
masses.

Figure \ref{fig:32Extrapolate} shows our values for \(a_\mu^{HVP}\)
plotted against Dashen mass, $\mu_q^D$.
Note that for ease of plotting, we have compressed the direction
relevant to the variation of $a_\mu^{HVP}$ with sea quark mass by shifting
all points to the physical sea quark masses \( \dlm_q = \dlm_q^{phys}
\).
The physical values for the valence quark masses are given by the red
(up), green (down) and blue (strange) vertical dashed lines.
The final value for $a_\mu^{HVP}$ is obtained by taking the
appropriate charge-weighted combination of all three quark flavour
contributions at their physical masses.
As this work is still preliminary, we refrain from quoting numbers at
this stage, however by comparing the results between the volumes it is
obvious that there are significant finite volume effects, particularly in
the \( \latSmall \) volume.
Given the analysis presented in Sec.~\ref{sec:fve}, this is not
surprising.

Finally, we note that the results from both volumes are described well
by the flavour-breaking expansions of Eq.~(\ref{eq:psMass}) and
that our use of partially-quenched valence quarks covering a large
range of masses and electric charges allows for contraints to be
placed on the various parameters.
In particular, we note the small difference in slopes between the red
(up quarks with charge $+2/3e$) and green (down/strange quarks with
charge $-1/3e$) curves which is a purely electromagnetic effect.
In future work we hope to improve the quality of the data and range of
ensembles available in order to isolate the contribution from the QED terms.

\section*{Acknowledgements}
The numerical configuration generation (using the BQCD lattice QCD
program \cite{Haar:2017ubh}) and data analysis (partly using the
Chroma software library \cite{Edwards:2004sx}) was carried out on the
IBM BlueGene/Q and HP Tesseract using DIRAC 2 resources (EPCC,
Edinburgh, UK), the IBM BlueGene/Q at NIC (J\"ulich, Germany), the
Cray XC40 at HLRN (The North-German Supercomputer Alliance), the NCI
National Facility in Canberra, Australia, and the iVEC facilities at
the Pawsey Supercomputing Centre.
These Australian resources are provided through the National
Computational Merit Allocation Scheme and the University of Adelaide
Partner Share supported by the Australian Government.
This work was supported in part through supercomputing resources
provided by the Phoenix HPC service at the University of Adelaide.
HP was supported by DFG Grant No. PE 2792/2-1,
PELR in part by the STFC under contract ST/G00062X/1 and
RDY and JMZ by the Australian Research
Council under grants FT120100821, FT100100005, and DP140103067.

\bibliographystyle{JHEP}
\bibliography{main}

\end{document}